\documentclass{jpp}
\usepackage{graphicx}
\usepackage{comment, color}

\usepackage[utf8]{inputenc}
\usepackage[T1]{fontenc}
\usepackage{amsmath}

\newcommand{\vect}[1]{\boldsymbol{#1}}
\newcommand{\Gsign}{ s_G }
\newcommand{\psisign}{ s_{\psi} }
\newcommand{\dR}{(R-R_0)}
\newcommand{\dz}{(z-z_0)}

\newcommand{\changed}[1]{{#1}}
\newcommand{\boozertor}{\varphi}

\shorttitle{Direct construction of optimized stellarator shapes. I.}
\shortauthor{M. Landreman and W. Sengupta}

\title{Direct construction of optimized stellarator shapes. I. Theory in cylindrical coordinates}

\author{Matt Landreman\aff{1}
  \corresp{\email{mattland@umd.edu}}
 \and Wrick Sengupta\aff{2}}

\affiliation{\aff{1}Institute for Research in Electronics and Applied Physics, University of Maryland, College Park MD 20742, USA
\aff{2}Courant Institute of Mathematical Sciences, New York University, New York NY 10012}

\begin{document}

\maketitle

\begin{abstract}
The confinement of guiding center trajectories in a stellarator is determined by the variation of the magnetic field strength $B$ in Boozer coordinates $(r, \theta, \boozertor)$, but $B(r,\theta,\boozertor)$ depends on the flux surface shape in a complicated way. 
Here we derive equations relating 
$B(r,\theta,\boozertor)$ in Boozer coordinates and the rotational transform
to the shape of flux surfaces in  cylindrical coordinates,
using an expansion in distance from the magnetic axis. A related expansion was done by Garren and Boozer [\emph{Phys. Fluids} B {\bf 3}, 2805 (1991)] based on the Frenet-Serret frame, which can be discontinuous anywhere the magnetic axis is straight, a situation that occurs in the interesting case of omnigenity with poloidally closed $B$ contours. Our calculation in contrast does not use the Frenet-Serret frame. 
The transformation between the Garren-Boozer approach and cylindrical coordinates is derived, and the two approaches are shown to be equivalent if the axis curvature does not vanish.
The expressions derived here help enable optimized plasma shapes to be constructed that can be provided as input to VMEC and other stellarator codes, or to generate initial configurations for conventional stellarator optimization.

\end{abstract}


\section{Introduction}

While stellarators offer the possibility of stable,
steady-state fusion power with minimal recirculating power
and immunity from disruptions, particle confinement in stellarators
is a challenge. In a general nonaxisymmetric magnetic field, even if magnetic surfaces exist, guiding center trajectories are not necessarily confined close to a magnetic surface in the absence of turbulence and collisions, as they are in perfect axisymmetry. However,
 confinement can be improved significantly by optimizing the shaping of the magnetic field. Guiding-center
trajectories are essentially determined by the strength
of the magnetic field $B$ in Boozer coordinates $(r,\theta,\boozertor)$, where $r$ labels magnetic surfaces, and
$\theta$ and $\boozertor$ are poloidal and toroidal angles \citep{BoozerCoordinates}.
If $B(r,\theta,\boozertor)$ has certain forms, such as quasisymmetry \citep{NuhrenbergZille} or omnigenity \citep{CaryShasharina,LandremanCatto},
the guiding center confinement would be as good as in  axisymmetry.
In principle, $B(r,\theta,\boozertor)$ is a function of the shapes of the magnetic surfaces through the equations of magnetohydrodynamic (MHD) equilibrium, but this functional relationship is complicated. Given a desired $B(r,\theta,\boozertor)$, it is not generally clear whether
a three-dimensional magnetic field $\vect{B}(\vect{r})$ exists with the desired field strength
and which solves the MHD equilibrium equations, much less what this solution $\vect{B}(\vect{r})$ is.

Previously, MHD equilibria with desirable $B(r,\theta,\boozertor)$ have been obtained using optimization \citep{NuhrenbergZille,NuhrenbergQA,Garabedian,NCSX}.
In this approach, an `off-the-shelf' optimization algorithm is applied to minimize an objective function representing the departure from the desired $B(r,\theta,\boozertor)$ (for instance, the summed squared amplitudes of symmetry-breaking terms in the Fourier series), as some shape parameters of a bounding magnetic surface are varied. For each function evaluation, a three-dimensional MHD equilibrium solution must be calculated numerically and then converted to Boozer coordinates.
While this approach has been successful, it has some shortcomings. 
Since there are multiple local minima, results depend on the initial condition, and one is never sure that 
all the interesting regions of parameter space have been found.
The optimization is computationally expensive, and little insight is gained as to the number of degrees of freedom in the problem. 

A complementary approach was taken by
\cite{GB1,GB2}.
Their work is commonly cited as a proof that perfectly 
quasisymmetric magnetic fields (apart from truly axisymmetric ones) do not exist,
but less well known is that their work contains a practical
procedure to directly construct MHD equilibria with desirable $B(r,\theta,\boozertor)$, generating ``optimized'' stellarators without optimization.
The Garren-Boozer analysis is based
upon an expansion in $r$, the effective distance from the magnetic axis; while it does not describe the outer region of a low-aspect-ratio device, it does describe some region sufficiently close to the axis of \emph{any} stellarator, even one with low aspect ratio.
(A complementary approach, based on expansion in departure from axisymmetry, was recently developed by \cite{PlunkHelander}.)
The present paper is the first in a series in which we
extend the Garren \& Boozer framework, to more fully understand the landscape of  stellarator shapes with good confinement, and to develop a practical tool for generating good initial conditions for conventional optimization.

In this first paper of the series, we derive the relationship between the shape of
the magnetic surfaces in cylindrical coordinates $(R,\phi,z)$
and $B$ in Boozer coordinates.
(More precisely, we consider surface shapes parameterized by $\{R(\theta,\phi),\,Z(\theta,\phi)\}$ using the Boozer poloidal angle $\theta$, so our representation is in a sense a hybrid one.)
While we use a similar $r$ expansion to Garren \& Boozer,
our calculation is different because theirs did not use cylindrical coordinates. Instead, Garren \& Boozer worked in the Frenet-Serret frame of the magnetic axis. The Frenet-Serret frame is an orthonormal basis $(\vect{t},\vect{n},\vect{b})$ satisfying the equations
\begin{align}
\label{eq:Frenet}
d{\vect{t}}/d\ell &= \kappa \vect{n}, \\ 
d{\vect{n}}/d\ell &= -\kappa \vect{t} + \tau\vect{b}, \nonumber \\ 
d{\vect{b}}/d\ell &= -\tau \vect{n}, \nonumber
\end{align}
where $\vect{t}=d\vect{r_0}/d\ell$, $\vect{r}_0$ is the position vector along the magnetic axis, and $\ell$ denotes the arclength along the curve. The vectors $\vect{t}$, $\vect{n}$, and $\vect{b}$ are called the tangent, normal, and binormal, $\kappa$ is the curvature, and $\tau$ is the torsion. Note that the opposite sign convention for torsion is used in \citep{GB1,GB2}.

There are two particular motivations for this paper.
First, we will (in Paper II of the series, \citep{PaperII}) generate plasma shapes as input for stellarator
physics codes that employ cylindrical coordinates,
specifically the VMEC code \citep{VMEC1983,VMEC1986}.
This can be done either using the equations for cylindrical coordinates derived in the present paper (section \ref{sec:direct}), or else
by solving Garren \& Boozer's
equations in the Frenet frame and mapping the results to cylindrical coordinates afterwards,
using a transformation that will be derived
in section \ref{sec:equivalence}.
By having these two approaches available, and showing
that the results are the same, we can be highly confident that the results are correct.
An analytic proof of the equivalence of the two
methods will be presented in this paper (section \ref{sec:equivalence}),
and numerical solutions will be presented in 
an accompanying Paper II \citep{PaperII}.
There, we will show that our approaches can generate
quasisymmetric flux surface shapes in $<$ 1 millisecond on a laptop -- 4 orders of magnitude faster than a single VMEC equilibrium calculation, much less a traditional optimization -- thus enabling high-resolution mapping of the landscape of possible quasisymmetric plasma shapes.

\begin{figure}
  \centering
  \includegraphics[width=4in]{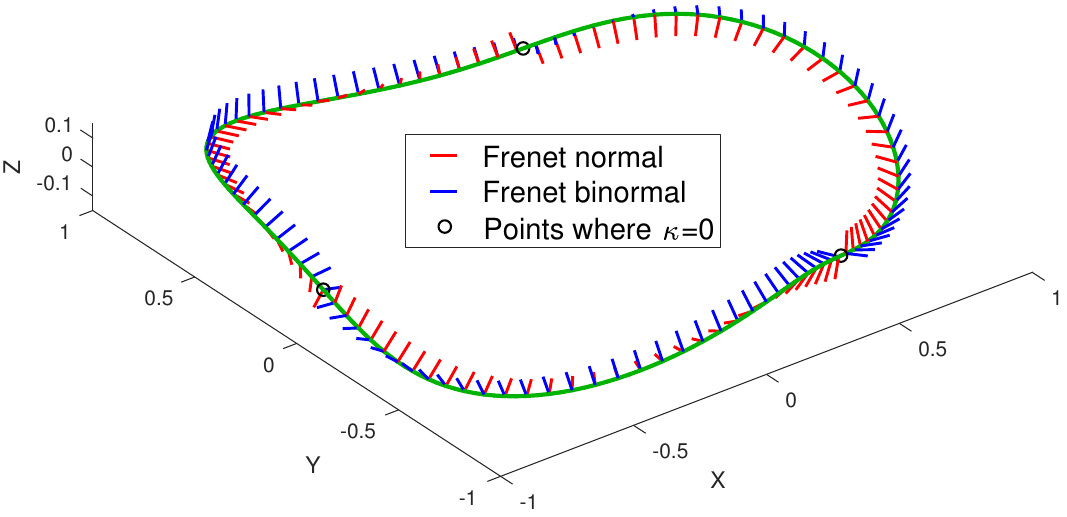}
  \caption{
  A smooth curve (green) for which the Frenet-Serret frame is discontinuous: 
  $R(\phi) = 1 + 0.1 \cos(3\phi)$, 
  $z(\phi) = 0.1 \sin(3\phi)$.
  }
\label{fig:discontinuity}
\end{figure}

Our second motivation in this paper is to modify Garren \& Boozer's analysis
to avoid the Frenet-Serret frame because this basis
can be pathological in certain situations of interest.
The Frenet-Serret frame is known to be problematic
if there are any points of vanishing curvature: even smooth curves
can have discontinuous Frenet-Serret basis vectors.
For instance, for the curve defined by $R(\phi) = 1 + R_c \cos(n\phi)$
and $z(\phi) = z_s \sin(n\phi)$, the curvature vanishes if
$R_c = 1/(n^2+1)$, and the Frenet basis is generally discontinuous at these points, as shown in figure \ref{fig:discontinuity}.
Where $\kappa=0$, the torsion is generally not well defined.
This situation of vanishing $\kappa$ is of particular interest because
it is \emph{necessary} for a desirable $B(r,\theta,\boozertor)$
optimization: omnigenity with poloidally closed $B$ contours
\citep{CaryShasharina,Subbotin,HelanderNuhrenberg,LandremanCatto} (sometimes called `quasi-isodynamic'.)
In this optimization, 
which yields good particle confinement at the same time as vanishing bootstrap current \citep{HelanderNuhrenberg},
the maximum of $B$ on each $r$ surface must be a constant-$\boozertor$ curve, so $\partial B/\partial\theta$ must vanish for all $\theta$ at these $\boozertor$ values. 
To see that this condition near the axis implies $\kappa=0$, consider that 
the pressure gradient $\nabla p$ vanishes on the magnetic axis, so it follows from the MHD equilibrium relation $(\nabla\times\vect{B})\times\vect{B}=0$ that
\begin{equation}
\label{eq:grad_perp_B}
\nabla_{\perp} B = \vect{B}\cdot\nabla(B^{-1}\vect{B}) = B\kappa\vect{n}.
\end{equation}
The condition $\partial B/\partial\theta$ on the maximum-$B$ curves near the axis implies $\nabla_\perp B=0$ there,
implying $\kappa=0$.
While one would have to grapple with discontinuities and ill-defined torsion to
apply the Frenet-Serret approach to construct omnigenous fields with poloidally closed $B$ contours, all quantities remain smooth in cylindrical coordinates.
Construction of omnigenous magnetic fields will
be considered in Paper III of this series.

The Frenet-Serret frame has also been used in another important stellarator calculation: Mercier’s result that rotational transform on the magnetic axis arises from a combination of axis torsion, rotating elongation, and current density \citep{Mercier,HelanderReview}. This result was also derived by \cite{GB1} as part of their quasisymmetry analysis, as their eq (77). Just as Garren \& Boozer’s quasisymmetry equation acquires singularities if the axis curvature ever vanishes, so does Mercier’s expression for the rotational transform, as it includes torsion explicitly. As part of our analysis, we will re-derive Mercier’s result in cylindrical coordinates, resulting in an expression that does not become singular if the axis curvature vanishes.

The main content of this paper begins in section \ref{sec:direct} with the calculation of the relationship
between $B(r,\theta,\boozertor)$ and flux surface shape   directly in cylindrical coordinates. 
The analogous results of the Garren-Boozer calculation
in the Frenet-Serret frame are then reviewed in section \ref{sec:Frenet}. The transformation between the two
coordinate systems is derived in section \ref{sec:equivalence_transformation}, and this transformation is used in the remainder of section \ref{sec:equivalence} to prove that the cylindrical and Frenet-Serret equations are equivalent, when the latter are valid.
Some reductions of the equations for the particular case
of quasisymmetry are discussed in section \ref{sec:quasisymmetry}, and we will conclude in section \ref{sec:conclusions}.


\section{Direct calculation in cylindrical coordinates}
\label{sec:direct}

We now present the calculation in which the field strength in Boozer coordinates is directly related to the magnetic surface shape in cylindrical coordinates.
Aside from the fact that we describe the magnetic surface shapes in cylindrical coordinates rather than by the projections along the Frenet-Serret vectors, our approach is similar in structure to the one in \cite{GB1}. The covariant and contravariant expressions for $\vect{B}$ in Boozer coordinates are equated, giving three independent equations. The square of either expression for $\vect{B}$  gives an additional equation for $B$. These four equations are then expanded in the distance $r$ from the magnetic axis.
Here we will carry out the expansion to sufficient order
that the first order quantities in $r$ are determined.


\subsection{Starting equations}

In any straight field line coordinates, including Boozer coordinates, the magnetic field can be written
\begin{align}
\vect{B} 
&= \nabla\psi \times\nabla\theta + \iota\nabla\boozertor\times\nabla\psi,
\label{eq:straight_field_lines}
\end{align}
where $2\pi\psi$ is the toroidal flux, $\iota$ is the rotational transform, and $\theta$ and $\boozertor$ are the poloidal and toroidal angles.
In the particular case of Boozer coordinates,
$\vect{B}$ can also be written
\begin{align}
\vect{B} &= \beta(\psi,\theta,\phi) \nabla\psi + I(\psi) \nabla \theta + G(\psi) \nabla\boozertor.
\label{eq:vacuum}
\end{align}
Here $I(\psi)$ is $\mu_0/(2\pi)$ times the toroidal current enclosed by the flux surface, and $G(\psi)$ is $\mu_0/(2\pi)$ times the poloidal current outside the flux surface. 
The Boozer toroidal angle $\boozertor$ differs from the cylindrical azimuthal angle $\phi$, and we will keep track of the difference, denoted $\nu$:
\begin{equation}
\boozertor = \phi + \nu.
\end{equation}
(By assuming this equation, our analysis will not pertain to certain unconventional configurations such as knots in which $\phi$ increases by an integer $>1$ multiple of $2\pi$ when $\boozertor$ increases by $2\pi$.)
We will consider the independent variables to be $(\psi,\theta,\phi)$.
From the product of (\ref{eq:straight_field_lines}) and (\ref{eq:vacuum}), the Jacobian of these coordinates is
\begin{equation}
\sqrt{g}
=\frac{1}{\nabla\psi \cdot\nabla\theta\times\nabla\phi} = \left( 1 + \frac{\partial\nu}{\partial\phi}\right) \frac{G+\iota I}{B^2}.
\end{equation}
We will assume $\partial\nu/\partial\phi>-1$ so this Jacobian remains nonzero. Physically, this assumption means the direction of $\vect{B}$ always points toward increasing $\phi$ or always points towards decreasing $\phi$, never reversing direction. This same assumption is made in the VMEC code \citep{VMEC1983}, and it is not restrictive in practice.

Using the dual relations
\changed{
\begin{align}
\frac{\partial\vect{r}}{\partial\psi} = \sqrt{g} \nabla\theta\times\nabla\phi,
\hspace{0.3in}
\nabla\psi = \frac{1}{\sqrt{g}} \frac{\partial\vect{r}}{\partial\theta}\times\frac{\partial\vect{r}}{\partial\phi},
\hspace{0.3in}\text{and cyclic permutations,}
\end{align}
where $\vect{r}$ is the position vector,
}
we can write (\ref{eq:straight_field_lines}) as
\begin{equation}
\vect{B} 
= 
 \frac{B^2}{G+\iota I}
\left[ \left( 1 + \frac{\partial\nu}{\partial\phi}\right) ^{-1}
\left( 1 - \iota \frac{\partial\nu}{\partial\theta}\right) \frac{\partial\vect{r}}{\partial\phi} 
+ \iota  \frac{\partial\vect{r}}{\partial\theta} 
\right],
\label{eq:B1}
\end{equation}
and write (\ref{eq:vacuum}) as
\begin{align}
\label{eq:B2}
\vect{B} = \frac{B^2}{G+\iota I}
& \left[
 \left( 1 + \frac{\partial\nu}{\partial\phi}\right) ^{-1}
\left(\beta+ G \frac{\partial\nu}{\partial\psi} \right)\frac{\partial\vect{r}}{\partial\theta} \times \frac{\partial\vect{r}}{\partial\phi} \right. \\
&+\left.  \left( 1 + \frac{\partial\nu}{\partial\phi}\right) ^{-1}
\left(I+G \frac{\partial\nu}{\partial\theta}\right) \frac{\partial\vect{r}}{\partial\phi}\times \frac{\partial\vect{r}}{\partial\psi}
+ G\frac{\partial\vect{r}}{\partial\psi} \times \frac{\partial\vect{r}}{\partial\theta}
\right].\nonumber
\end{align}
The derivatives of $\vect{r}(\psi,\theta,\phi)=R \vect{e}_R
+z\vect{e}_z$ can be evaluated using $d \vect{e}_R/d\phi=\vect{e}_\phi$,
where $(\vect{e}_R,\vect{e}_\phi,\vect{e}_z)$ are cylindrical unit basis vectors.
Equating the three cylindrical components of (\ref{eq:B1}) and (\ref{eq:B2}), we obtain
\begin{equation}
\frac{r \bar{B}}{R}
\left[ 
\left( 1 - \iota \frac{\partial\nu}{\partial\theta} \right)
 \frac{\partial R}{\partial\phi} 
 + \iota   \left( 1+ \frac{\partial\nu}{\partial\phi}\right) \frac{\partial R}{\partial\theta} \right]
=
\left( I+G\frac{\partial\nu}{\partial\theta}\right)  \frac{\partial z}{\partial r} 
-\left(\beta r \bar{B} + G \frac{\partial\nu}{\partial r} \right)\frac{\partial z}{\partial\theta},
\label{eq:BR}
\end{equation}
\begin{align}
&\frac{r \bar{B}}{GR}
\left\{
\left( 1 - \iota \frac{\partial\nu}{\partial\theta}\right)
\left[ R^2 + \left(  \frac{\partial R}{\partial\phi}\right)^2 +\left(\frac{\partial z}{\partial\phi}\right)^2\right]
+\iota \left(1+\frac{\partial\nu}{\partial\phi}\right) 
\left( \frac{\partial R}{\partial\theta} \frac{\partial R}{\partial\phi} + \frac{\partial z}{\partial\theta} \frac{\partial z}{\partial\phi}
\right) \right\} \nonumber\\
&=
\left( \frac{\partial z}{\partial  r} \frac{\partial R}{\partial\theta}
- \frac{\partial R}{\partial r} \frac{\partial z}{\partial\theta}  \right)  \left( 1 + \frac{\partial\nu}{\partial\phi}\right),
\label{eq:Bphi2}
\end{align}
\begin{equation}
 \frac{r \bar{B}}{R}
\left[ 
\left( 1 - \iota \frac{\partial\nu}{\partial\theta}\right)
 \frac{\partial z}{\partial\phi} 
 + \iota  \left( 1+ \frac{\partial\nu}{\partial\phi}\right) \frac{\partial z}{\partial\theta} \right]
 =
\left( \beta r \bar{B} + G\frac{\partial\nu}{\partial r}\right)  \frac{\partial R}{\partial\theta} 
-\left(I + G \frac{\partial\nu}{\partial\theta}\right) \frac{\partial R}{\partial r}.
\label{eq:Bz}
\end{equation}
To get (\ref{eq:Bphi2}) we have added 
(\ref{eq:BR}) times $\partial R/\partial\phi$ and (\ref{eq:Bz}) times $\partial z/\partial\phi$ to the $\vect{e}_\phi$ components.
In these expressions, we have changed the flux surface label coordinate from $\psi$ to the effective minor radius $r(\psi)$ defined by $2 \pi \psi = \pi r^2 \bar{B}$,
where $\bar{B}$ is an arbitrary reference magnitude of magnetic field.
(Since $\psi$ can be negative, $\bar{B}$ may be negative.)
Also, a relation for $B$ can be obtained by squaring (\ref{eq:B1}):
\begin{align}
\frac{(G+\iota I)^2}{B^2}
\left( 1 + \frac{\partial\nu}{\partial\phi}\right) ^2
=&
\left[ 
\left( 1 - \iota \frac{\partial\nu}{\partial\theta}\right)
 \frac{\partial R}{\partial\phi} 
 + \iota  \left( 1 + \frac{\partial\nu}{\partial\phi}\right) \frac{\partial R}{\partial\theta} \right] ^2 
 +
\left( 1 - \iota \frac{\partial\nu}{\partial\theta}\right)^2
R^2\nonumber \\
&+ 
\left[ 
\left( 1 - \iota \frac{\partial\nu}{\partial\theta}\right)
 \frac{\partial z}{\partial\phi} 
 + \iota  \left( 1 + \frac{\partial\nu}{\partial\phi}\right) \frac{\partial z}{\partial\theta} \right]^2.
\label{eq:modB}
\end{align}

Equations (\ref{eq:BR}) - (\ref{eq:modB}) are the basis of the remainder of the analysis,
in which these equations will be systematically expanded.


\subsection{Expansion about the magnetic axis}

\changed{We take the magnetic axis to be described by its cylindrical coordinates $R_0(\phi)$ and $z_0(\phi)$.
Regularity considerations near the axis imply we can write the cylindrical coordinate $R(r,\theta,\phi)$ 
for a general point near the axis in the form of an expansion}
\begin{align}
\label{eq:RExpansion}
R(r,\theta,\phi) &= R_0(\phi) + r R_{1} (\theta,\phi) + r^2 R_{2} (\theta,\phi) + \ldots
\end{align}
\changed{where}
\begin{align}
R_1(\theta,\phi) &= R_{1c} (\phi) \cos\theta + R_{1s} (\phi) \sin\theta, \label{eq:R1} \\
R_2(\theta,\phi) &= R_{2c} (\phi) \cos 2\theta + R_{2s} (\phi) \sin 2\theta + R_{20}(\phi).
\label{eq:R2Expansion}
\end{align}
\changed{Expansions of the same form are made for $z$, $\nu$ and $B$:
\begin{align}
z = &z_0(\phi) + r[z_{1c}(\phi)\cos\theta + z_{1s}(\phi)\sin\theta]
+r^2[z_{20}(\phi) + z_{2c}(\phi)\cos 2\theta + z_{2s}(\phi)\sin 2\theta]+\ldots
\nonumber \\
\nu = &\nu_0(\phi) + r[\nu_{1c}(\phi)\cos\theta + \nu_{1s}(\phi)\sin\theta]
+r^2[\nu_{20}(\phi) + \nu_{2c}(\phi)\cos 2\theta + \nu_{2s}(\phi)\sin 2\theta]+\ldots
\nonumber \\
B = &B_0(\phi) + r[B_{1c}(\phi)\cos\theta + B_{1s}(\phi)\sin\theta]
\nonumber \\
&+r^2[B_{20}(\phi) + B_{2c}(\phi)\cos 2\theta + B_{2s}(\phi)\sin 2\theta]+\ldots.
\end{align}
These expansions are justified in appendix \ref{sec:regularity}.}
We also have
\begin{align}
G(r) & = G_0 + r^2 G_2 + \ldots, \\
I(r) & = r^2 I_2 + \ldots, \\
\beta(r,\theta,\phi) & = \beta_0(\phi) + r \beta_1(\theta,\phi) + \ldots, \\
\iota(r) & = \iota_0 + \ldots.
\end{align}
Using these expansions, we proceed to systematically
consider the terms of each order in (\ref{eq:BR})
- (\ref{eq:modB}).


\subsection{Magnitude of $B$: zeroth order}

We first consider the $O(r^0)$ terms in (\ref{eq:modB}). These terms give
\begin{equation}
\nu'_0
= -1 +\Gsign\ell' B_0 / G_0,
\label{eq:nu0Solve}
\end{equation}
where $\Gsign = \pm 1$, primes denote $d/d\phi$, 
and $ \ell'>0$ is the differential length of the magnetic axis:
\begin{equation}
\ell'
=
\sqrt{
R_0^2+ 
(  R'_0) ^2 +
( z'_0)^2}.
\end{equation}
Integrating (\ref{eq:nu0Solve}) in $\phi$, 
\begin{equation}
G_0
=
\frac{\Gsign}{2\pi} 
\int_0^{2\pi}d\phi\; B_0
 \ell'.
\label{eq:B0OverG}
\end{equation}
Thus, $\Gsign$ is the sign of $G_0$,
$+1$ if $\vect{B}$ points in the direction of increasing $\phi$ and $-1$ otherwise.
Equations (\ref{eq:nu0Solve})-(\ref{eq:B0OverG}) allow us to eliminate $\nu_0$ and $G_0$ in favor of
$R_0$, $z_0$, and $B_0$.

\subsection{Equating representations of the field: first order}

Next, 
the leading-order terms in the $r$ expansion of 
(\ref{eq:Bphi2})
are $O(r^1)$, giving
\begin{equation}
\frac{\bar{B}}{G_0 R_0}
\left( \ell'\right)^2
=
\left( R_{1s}  z_{1c} - R_{1c} z_{1s} \right)  \left( 1 + \nu'_0\right).
\label{eq:1st_order_Bphi_constraint}
\end{equation}
We can eliminate $\nu_0$ in this equation using (\ref{eq:nu0Solve}) to obtain
\begin{equation}
\frac{\Gsign \bar{B} \ell'}{R_0 B_0} 
=
R_{1s}  z_{1c} - R_{1c} z_{1s}.
\label{eq:const_B_on_axis}
\end{equation}
This equation, which is analogous to (53) in \cite{GB1}, 
expresses the fact that the toroidal flux 
within the magnetic surface $r$ should be 
$2\pi\psi=\pi r^2 \bar{B}$. To see this, consider that the toroidal
field on the magnetic axis is $\vect{B}\cdot\vect{e}_\phi = \Gsign B_0 \vect{t}\cdot\vect{e}_\phi = B_0 R_0/(\Gsign \ell')$, and as shown in appendix \ref{sec:ellipse},
the area of the flux surface in the constant-$\phi$ plane is
$\pi r^2$ times the right hand side of (\ref{eq:const_B_on_axis}).

Similarly, the leading terms in (\ref{eq:BR}) and (\ref{eq:Bz})
are $O(r^1)$ and give
\begin{align}
\frac{\bar{B} R'_0}{G_0 R_0}
&= \nu_{1s}  z_{1c} - \nu_{1c} z_{1s},
\label{eq:BR1}
\\
\frac{\bar{B} z'_0}{G_0 R_0}
&= \nu_{1c}  R_{1s} - \nu_{1s} R_{1c}.
\label{eq:Bz1}
\end{align}
Solving for $\nu_{1c}$ and $\nu_{1s}$ and applying (\ref{eq:const_B_on_axis}), we find
\begin{equation}
\nu_1 = \frac{B_0}{|G_0| \ell'}  \left( R_1 R'_0+ z_1 z'_0\right).
\label{eq:nu1}
\end{equation}


\subsection{Magnitude of $B$: first order}

Another pair of equations is obtained from the $O(r^1)$ terms in (\ref{eq:modB}). These terms can be found by applying $\partial/\partial r$ to (\ref{eq:modB}) and evaluating the result at $r \to 0$.
We find
\begin{align}
&-\frac{G_0^2 B_1}{B_0^3}\left( 1 + \nu'_0\right)^2 + \frac{G_0^2}{B_0^2} \left( 1 + \nu'_0\right) \frac{\partial \nu_1}{\partial\phi}
\nonumber \\
&=
R'_0\left[ -\iota_0 \frac{\partial \nu_1}{\partial\theta} R'_0 + \frac{\partial R_1}{\partial\phi} 
+ \iota_0\left(1 + \nu'_0\right)\frac{\partial R_1}{\partial\theta}\right]
+ R_0 R_1 - \iota_0 \frac{\partial \nu_1}{\partial\theta} R_0^2 \nonumber \\
& \;\;\;+
z'_0\left[ -\iota_0 \frac{\partial \nu_1}{\partial\theta} z'_0 + \frac{\partial z_1}{\partial\phi} 
+ \iota_0\left(1 + \nu'_0\right)\frac{\partial z_1}{\partial\theta}\right].
\label{eq:modB1}
\end{align}
In this equation, the terms that include a factor of $\iota_0$ can be written
\begin{equation}
\iota_0 \frac{\partial}{\partial\theta} \left[
-\left(\ell'\right)^2 \nu_1 + \left( 1 + \nu'_0\right) \left( R_1 R'_0 + z_1 z'_0\right)\right],
\label{eq:iota_terms}
\end{equation}
which can be seen to vanish in light of (\ref{eq:nu1}) and (\ref{eq:nu0Solve}). 
Eliminating $\nu_0$ and $\nu_1$ in the remaining terms
using (\ref{eq:nu0Solve}) and (\ref{eq:nu1}),
one finds
\begin{equation}
B_1 / B_0 = K_R R_{1} + K_z z_{1},
\label{eq:B1s}
\end{equation}
where
\begin{align}
K_R
&=  -\left( \ell'\right)^{-4}
\left( R_0 R'_0 + R'_0 R''_0 +z'_0 z''_0\right)
 R'_0
+ \left(  \ell'\right)^{-2} 
\left( R''_0 - R_0 
+R'_0 B'_0/B_0 \right), \label{eq:KR} \\
K_z
&=  -\left( \ell'\right)^{-4}
\left( R_0 R'_0 + R'_0 R''_0 + z'_0 z''_0\right)
 z'_0
+ \left(  \ell'\right)^{-2} 
\left( z''_0
+z'_0 B'_0/B_0 \right).
\label{eq:Kz}
\end{align}
Noting from the first line of (\ref{eq:Frenet}) that $\kappa \vect{n} \ell' = \vect{t}' = [(\ell')^{-1}\vect{r}'_0]'$, and evaluating the result in cylindrical coordinates, it can be seen that equivalent expressions to (\ref{eq:KR})-(\ref{eq:Kz}) are
\begin{align}
K_R = \kappa \vect{n}\cdot\vect{e}_R 
+ (\ell')^{-2} R'_0 B'_0 / B_0, 
\hspace{0.5in}
K_z = \kappa \vect{n}\cdot\vect{e}_z 
+ (\ell')^{-2} z'_0 B'_0 / B_0.
\end{align}
Note that the $\sin\theta$ and $\cos\theta$ components 
of $B_1$, $R_1$, and $z_1$ each satisfy
(\ref{eq:B1s}) separately.
Equations (\ref{eq:B1s})-(\ref{eq:Kz}) are analogous
to (70) in \cite{GB1}.
These equations reflect (\ref{eq:grad_perp_B}).
In the limit of a circular magnetic axis, $R'_0=0$ and $z'_0=0$, (\ref{eq:B1s})-(\ref{eq:Kz}) reduce to $B_1 / B_0 = - R_1 / R_0$, reflecting the expected $B \propto 1/R$ variation.


\subsection{Equating representations of the field: second order}

The highest order terms in the $r$ expansion we will consider 
are the $O(r^2)$ terms in (\ref{eq:BR})-(\ref{eq:Bz}).
The expressions at this order become rather lengthy and so details
are left to appendix \ref{sec:2nd_order}.
At $O(r^2)$, the three equations (\ref{eq:BR})-(\ref{eq:Bz})
each have a $\sin\theta$ and $\cos\theta$ component, so
there are six independent equations.
Although nine second order quantities ($R_{2s}$, $R_{2c}$, $R_{20}$, and similar $\nu$ and $z$ terms) appear, 
they only enter through five linearly independent
combinations. Therefore
the second order quantities can be annihilated by forming
a certain linear combination of the six equations, (\ref{eq:annihilator}).
What remains is an equation relating zeroth and first order quantities:
\begin{equation}
\iota_0  V - T = 0,
\label{eq:iota}
\end{equation}
where
\begin{align}
\label{eq:numerator}
T = 
\frac{|G_0|}{ (\ell')^3 B_0}
 &\left[
R_0^2 \left( R_{1c} R'_{1s}-R_{1s}R'_{1c}+z_{1c}z'_{1s}-z_{1s}z'_{1c}\right) \right. \\
&+\left(R_{1c}z_{1s}-R_{1s}z_{1c}\right) 
\left(R'_0 z''_0 
+2R_0 z'_0
-z'_0 R''_0 \right)\nonumber\\
&+\left(z_{1c}z'_{1s}-z_{1s}z'_{1c}\right)\left(R'_0\right)^2
+\left(R_{1c}R'_{1s}-R_{1s}R'_{1c}\right)\left( z'_0 \right)^2  \nonumber \\
&+\left.\left(R_{1s}z'_{1c}-z_{1c}R'_{1s}+z_{1s}R'_{1c}-R_{1c}z'_{1s}\right) R'_0 z'_0\right]
+\frac{2 G_0 I_2}{B_0^2}  \nonumber
\end{align}
and
\begin{align}
\label{eq:denominator}
V= &
\frac{1}{(\ell')^2}
\left[
R_0^2 \left( R_{1c}^2 + R_{1s}^2 + z_{1c}^2 + z_{1s}^2 \right)
+\left(R'_0\right)^2 \left(z_{1c}^2+z_{1s}^2\right) \right.
\\
&\left.  \hspace{0.5in} -2R'_0z'_0\left(R_{1c}z_{1c}+R_{1s}z_{1s}\right)
+\left(z'_0\right)^2\left(R_{1c}^2+R_{1s}^2\right)\right]. \nonumber 
\end{align}
Our (\ref{eq:iota})-(\ref{eq:denominator}) play an analogous role to (63) and (67) in \cite{GB1}.
Note that  (\ref{eq:iota}) can be integrated to give $\iota = \left(\oint w\, d\phi\right)^{-1} \oint (wT/V)d\phi$ for any $w(\phi)$, analogous to Garren \& Boozer's (77).
Encoded in these equations is the classic result by
Mercier \citep{Mercier, HelanderReview}: rotational transform on the magnetic axis
arises due to axis torsion, rotating elongation,
and toroidal current. Indeed, in Paper II we will compute $\iota_0$ numerically by solving (\ref{eq:iota})-(\ref{eq:denominator}) or its Frenet-Serret analogue. The toroidal current contribution to $\iota_0$ is
the $I_2$ term in $T$, while the axis torsion and rotating elongation contributions are evidently contained in the remaining terms. 
Interestingly, while the torsion in Mercier's expression involves the third derivative of the axis shape, the highest derivative of the axis shape appearing in (\ref{eq:iota})-(\ref{eq:denominator}) is the second.
If there are any points where the axis curvature vanishes,
the torsion becomes ill-defined so Mercier's expression for $\iota$ (which explicitly depends on $\tau$) becomes awkward;  (\ref{eq:iota}) has no such problem.

\changed{Another perspective on rotational transform and torsion in cases with vanishing curvature (without effects of elongation) has been discussed by \cite{Pfefferle}.}


\section{Frenet-Serret approach}
\label{sec:Frenet}

The analogous calculation using the Frenet-Serret frame
is clearly explained in \cite{GB1,GB2}, so we will not repeat it here,
only quote the main results.
The position vector is written 
\begin{equation}
\vect{r}(r,\theta,\boozertor)=\vect{r}_0(\boozertor) + X(r,\theta,\boozertor) \vect{n}(\boozertor) + Y(r,\theta,\boozertor) \vect{b}(\boozertor) + Z(r,\theta,\boozertor) \vect{t}(\boozertor),
\label{eq:GBr}
\end{equation}
where $\vect{r}_0$, $\vect{n}$, $\vect{b}$, and $\vect{t}$ refer to the magnetic axis. The quantities $X$, $Y$, and $Z$ are expanded 
similarly to (\ref{eq:RExpansion})-(\ref{eq:R2Expansion}) but with $\phi \to \boozertor$:
\begin{align}
X(r,\theta,\boozertor) &= r X_{1} (\theta,\boozertor) + r^2 X_{2} (\theta,\boozertor) + \ldots
\end{align}
where regularity requires
\begin{align}
X_1(\theta,\boozertor) &= X_{1c} (\boozertor) \cos\theta + X_{1s} (\boozertor) \sin\theta,
\\
X_2(\theta,\boozertor) &= X_{2c} (\boozertor) \cos 2\theta + X_{2s} (\boozertor) \sin 2\theta + X_{20}(\boozertor),
\end{align}
and analogous expansions are made for $Y$ and $Z$. The expansion of $B$ is written in terms of $\boozertor$ rather than $\phi$, so
\begin{equation}
\label{eq:B1_Frenet}
B(r,\phi,\boozertor) = \hat{B}_0(\boozertor) + r \hat{B}_1(\theta,\boozertor) + \ldots
\end{equation}
where $\hat{B}_1(\theta,\boozertor) = \hat{B}_{1s}(\boozertor)\sin\theta + \hat{B}_{1c}(\boozertor)\cos\theta$.

Instead of (\ref{eq:nu0Solve}), one obtains
$G_0 = \Gsign B_0 \, d\ell/d\boozertor$.
Instead of (\ref{eq:const_B_on_axis}), one finds $Z_1=0$ and
\begin{align}
X_{1c} Y_{1s} - X_{1s} Y_{1c} =  \Gsign \bar{B} / B_0 .
\label{eq:GBNonlinear1}
\end{align}
Noting from appendix \ref{sec:ellipse} that the left hand side of this equation
is the cross-sectional area of the flux surface in a plane perpendicular to the on-axis $\vect{B}$,
(\ref{eq:GBNonlinear1}) represents the fact that the toroidal flux 
inside the flux surface is $\pi r^2 \bar{B}$.
Instead of (\ref{eq:B1s}), one finds
\begin{equation}
\hat{B}_1 / \hat{B}_0 = \kappa X_1,
\label{eq:GB_B1}
\end{equation}
where this equation holds separately for $\sin\theta$ and $\cos\theta$
components.
Instead of (\ref{eq:iota})-(\ref{eq:denominator}), Garren \& Boozer obtain
\begin{equation}
\iota_0 V^{FS} - T^{FS} = 0,
\label{eq:GB2nd_order}
\end{equation}
where
\begin{align}
V^{FS} =  X_{1s}^2 + X_{1c}^2 + Y_{1s}^2 + Y_{1c}^2 
\label{eq:GBDenominator}
\end{align}
and
\begin{align}
T^{FS} =  X_{1c} \frac{d  X_{1s}}{d \boozertor} - X_{1s} \frac{d  X_{1c}}{d \boozertor}  
+  Y_{1c} \frac{d  Y_{1s}}{d \boozertor} - Y_{1s} \frac{d  Y_{1c}}{d \boozertor} 
+2 \left( \frac{I_2}{\bar{B}} - \tau \right) \frac{ G_0\bar{B}}{ B_0^2}.
\label{eq:GBNumerator}
\end{align}
These equations correspond to (63) in \cite{GB1}, but with an extra $I_2$ term since a vacuum field was assumed
in that work. The fact that a $2$ appears in the $\tau$ term here whereas a 4 appears in \cite{GB1} is due to the normalization used in the latter, and $\tau$ enters with the opposite sign due to the opposite sign convention for torsion.

Combining the above equations to eliminate unknowns,
the system can be reduced to a single equation.
To this end, we introduce a variable $\sigma(\boozertor)$ related to the flux surface shape, defined by
\begin{equation}
\Gsign \bar{B} \kappa \sigma = \hat{B}_{1s} Y_{1s} + \hat{B}_{1c} Y_{1c} .
\end{equation}
From this definition and (\ref{eq:GBNonlinear1})-(\ref{eq:GB_B1}), 
\begin{align}
\label{eq:W_to_Y}
Y_{1s} &= \frac{\Gsign \bar{B} \kappa}{\hat{B}_{1s}^2 + \hat{B}_{1c}^2}(\hat{B}_{1c} + \hat{B}_{1s} \sigma), \\
Y_{1c} &= \frac{\Gsign \bar{B} \kappa}{\hat{B}_{1s}^2 + \hat{B}_{1c}^2}(-\hat{B}_{1s} + \hat{B}_{1c} \sigma). \nonumber
\end{align}
Substituting these results and (\ref{eq:GB_B1}) into (\ref{eq:GB2nd_order}),
we obtain
\begin{align}
\label{eq:sigma_general}
\frac{d\sigma}{d\boozertor}
+\left[ 
\frac{(\hat{B}_{1s}^2+\hat{B}_{1c}^2)^2}{B_0^2 \bar{B}^2 \kappa^4}
+1+\sigma^2 \right]
\left[ \iota_0 + 
\frac{1}{\hat{B}_{1s}^2 + \hat{B}_{1c}^2}
\left( \hat{B}_{1s}\frac{d\hat{B}_{1c}}{d\boozertor} - \hat{B}_{1c}\frac{d\hat{B}_{1s}}{d\boozertor}\right) \right] &\\
-2 \left(\frac{I_2}{\bar{B}} - \tau \right) \frac{ G_0(\hat{B}_{1s}^2+\hat{B}_{1c}^2)}{ \bar{B}B_0^2 \kappa^2}
&=0.\nonumber
\end{align}
Considering $\kappa$, $\tau$, $I_2$, $B_0$, $\hat{B}_{1s}$, 
and $\hat{B}_{1c}$ to be known, this result is a first-order
nonlinear ODE for $\sigma$. Once $\sigma$ is obtained, $Y_{1s}$ and $Y_{1c}$
can be found from (\ref{eq:W_to_Y}), and $X_{1s}$ and $X_{1c}$ are
known from (\ref{eq:GB_B1}), so the flux surface shape can be reconstructed
from (\ref{eq:GBr}).



\section{Equivalence of the two approaches}
\label{sec:equivalence}


\subsection{Relating representations of the surface shape}
\label{sec:equivalence_transformation}

Let us now prove that if the curvature of the magnetic
axis does not vanish, the Frenet-Serret approach
and the direct calculation in cylindrical coordinates
are equivalent, as they should be.
To begin, we must relate $X_1$ and $Y_1$ to $R_1$ and $z_1$. This can be done by equating the position vector in the two
approaches, expanding (\ref{eq:GBr}) using
$\boozertor(r,\theta,\phi) = \boozertor_0(\phi) + r \nu_1(\theta,\phi) + O(r^2)$ where
$\boozertor_0(\phi) = \phi + \nu_0(\phi)$:
\begin{align}
\label{eq:equating_position_vectors}
&\left[ R_0(\phi) + r R_1(\theta,\phi) \right]\vect{e}_R(\phi)
+\left[ z_0(\phi) + r z_1(\theta,\phi)\right]\vect{e}_z + O(r^2) \\
&=\vect{r}_0(\boozertor_0) + r \nu_1(\theta,\phi) 
\,d\vect{r}_0/d\boozertor_0
+r X_1(\theta,\boozertor_0) \vect{n}(\boozertor_0)
+r Y_1(\theta,\boozertor_0) \vect{b}(\boozertor_0)
+O(r^2). \nonumber
\end{align}
Equating the $O(r^0)$ terms gives $\vect{r}_0(\boozertor_0)
= R_0(\phi) \vect{e}_R(\phi) + z_0(\phi) \vect{e}_z$.
Then applying $\vect{n}(\boozertor_0)\cdot(\ldots)$ and $\vect{b}(\boozertor_0)\cdot(\ldots)$ to the $O(r)$ terms in (\ref{eq:equating_position_vectors}), we obtain two equations
that can be represented
\begin{equation}
\begin{pmatrix} X_1 \\ Y_1 \end{pmatrix}
= \begin{pmatrix}
n_R & n_z \\
b_R & b_z
\end{pmatrix}
\begin{pmatrix} R_1 \\ z_1 \end{pmatrix}.
\label{eq:original_transformation}
\end{equation}
Here and for the rest of this section, 
$n_R = \vect{n}(\boozertor_0)\cdot\vect{e}_R(\phi)$,
$b_R = \vect{b}(\boozertor_0)\cdot\vect{e}_R(\phi)$,
analogous expressions hold for $n_z$ and $b_z$,
and $X_1$ and $Y_1$ are understood to be evaluated at $\boozertor_0$.
(The $\vect{t}(\boozertor_0)\cdot(\ldots)$ component of (\ref{eq:equating_position_vectors}) yields (\ref{eq:nu1}).)

Noting the components of the tangent vector in cylindrical coordinates,
\begin{equation}
t_R = \vect{t}\cdot\vect{e}_R = R'_0 / \ell', 
\hspace{0.4in}
t_\phi = \vect{t}\cdot\vect{e}_\phi = R_0 / \ell', 
\hspace{0.4in}
t_z = \vect{t}\cdot\vect{e}_z = z'_0 / \ell',
\end{equation}
the determinant of the matrix in (\ref{eq:original_transformation}) is
\begin{equation}
n_R b_z - b_R n_z
=-\vect{n}\times\vect{b}\cdot\vect{e}_\phi
=-\vect{t}\cdot\vect{e}_\phi = -R_0 / \ell'.
\label{eq:transformation_determinant}
\end{equation}
Hence the inverse transformation is
\begin{equation}
\begin{pmatrix} R_1 \\ z_1 \end{pmatrix}
= 
\frac{\ell'}{R_0}\begin{pmatrix}
-b_z & n_z \\
b_R & -n_R
\end{pmatrix}
\begin{pmatrix} X_1 \\ Y_1 \end{pmatrix}.
\label{eq:transformation}
\end{equation}
This relation enables the solution of the quasisymmetry
equations in the Frenet-Serret basis to be mapped
to cylindrical coordinates.
Note that by applying (\ref{eq:transformation}) and (\ref{eq:transformation_determinant}) to 
(\ref{eq:const_B_on_axis}),
we obtain (\ref{eq:GBNonlinear1}),
and so these equations from the Frenet-Serret and
cylindrical coordinates analyses are consistent.


\subsection{Equivalence of the $B_1$ equations}

Next let us show that (\ref{eq:B1s}) and (\ref{eq:GB_B1}) are equivalent. 
Expanding (\ref{eq:B1_Frenet}) about $\boozertor \approx \boozertor_0$,
and equating the result to the $B$ analogue of (\ref{eq:RExpansion}),
we obtain
\begin{align}
B_0(\phi) + r B_1(\theta,\phi) + O(r^2)
=\hat{B}_0(\boozertor_0) 
+ r \nu_1(\theta,\phi) 
\,d\hat{B}_0/d\boozertor_0
+r \hat{B}_1(\theta,\boozertor_0)+O(r^2).
\label{eq:equating_B}
\end{align}
The $O(r^0)$ terms give $B_0(\phi)=\hat{B}_0(\boozertor_0)$,
which upon differentiation gives
\begin{equation}
B'_0(\phi) = \left[1 + \nu'_0(\phi)\right]
d\hat{B}_0/d\boozertor_0.
\end{equation}
Combining this result with (\ref{eq:nu0Solve}), (\ref{eq:nu1}),
and the $O(r^1)$ terms of (\ref{eq:equating_B}),
we find 
\begin{align}
\hat{B}_1(\theta,\boozertor_0) = B_1 - B'_0 (\ell ')^{-2}(R'_0 R_1 + z'_0 z_1).
\label{eq:B_cyl_to_Boozer}
\end{align}
Then using the top row of (\ref{eq:original_transformation}), (\ref{eq:GB_B1}) and  (\ref{eq:B1s}) are equivalent.
Note that using (\ref{eq:B_cyl_to_Boozer}), (\ref{eq:B1s}) can be written in terms of $\hat{B}_1$ rather than $B_1$, yielding a relation between the flux surface shape in cylindrical coordinates and the field strength in Boozer coordinates:
\begin{equation}
\hat{B}_1(\theta,\boozertor_0) / \hat{B}_0(\boozertor_0)= (n_R R_{1} + n_z z_{1}) \kappa.
\label{eq:B1_mixedCoords}
\end{equation}


\subsection{Equivalence of the $\iota_0$ equations}

Finally, let us show that equations (\ref{eq:iota})-(\ref{eq:denominator}),
which determine $\iota_0$ in cylindrical coordinates,
can be independently derived from the analogous Frenet-Serret equations (\ref{eq:GB2nd_order})-(\ref{eq:GBNumerator})
by applying the transformation (\ref{eq:original_transformation}).
We first note the following relations between components of the normal
and binormal vectors:
\begin{align}
\label{eq:identities}
n_R^2 + b_R^2 
&=\left[(\vect{t}\vect{t}+\vect{n}\vect{n}+\vect{b}\vect{b})\cdot\vect{e}_R\right]^2-t_R^2 = 1-t_R^2
= \frac{R_0^2 + (z'_0)^2}{(\ell')^2},
\\
n_z^2 + b_z^2 
&=\left[(\vect{t}\vect{t}+\vect{n}\vect{n}+\vect{b}\vect{b})\cdot\vect{e}_z\right]^2-t_z^2 = 1-t_z^2
= \frac{R_0^2 + (R'_0)^2}{(\ell')^2}, \nonumber\\
n_R n_z + b_R b_z 
&= \vect{e}_R\cdot(\vect{t}\vect{t}+\vect{n}\vect{n}+\vect{b}\vect{b})\cdot\vect{e}_z-t_R t_z = \vect{e}_R\cdot\vect{e}_z-t_R t_z
=-t_R t_z = -R'_0 z'_0 / (\ell')^2.
\nonumber
\end{align}
Using these results and (\ref{eq:transformation}), then
\begin{align}
X_{1s}^2 + Y_{1s}^2 
&=(\ell')^{-2} \left[R_0^2(R_{1s}^2+z_{1s}^2) + (z'_0)^2R_{1s}^2
-2R'_0 z'_0 R_{1s} z_{1s} + (R'_0)^2 z_{1s}^2 \right]. \nonumber
\end{align}
An analogous expression holds for the subscript-$1c$ ($\cos\theta$) terms.
Thus, it can be seen that $ V^{FS} = V$.

It remains to show $T^{FS} = T$. To show this equivalence we first apply (\ref{eq:original_transformation}) and then (\ref{eq:identities}) to the first four terms of $T^{FS}$, giving
\begin{align}
T^{FS} 
&=\frac{|G_0|}{B_0 (\ell')^3}[
R_0^2(R_{1c}R'_{1s} - R_{1s}R_{1c}' + z_{1c}z'_{1s} - z_{1s}z'_{1c}) \\
&+\left(z_{1c}z'_{1s}-z_{1s}z'_{1c}\right)\left(R'_0\right)^2 
+\left(R_{1c}R'_{1s}-R_{1s}R'_{1c}\right)\left( z'_0 \right)^2 \nonumber \\  
&+\left.\left(R_{1s}z'_{1c}-z_{1c}R'_{1s}+z_{1s}R'_{1c}-R_{1c}z'_{1s}\right) R'_0 z'_0\right]
+ \frac{2 I_2 G_0}{B_0^2}
+\hat{T},  \nonumber
\end{align}
where
\begin{align}
\hat{T}= \frac{|G_0|}{B_0 \ell'}
(R_{1s} z_{1c} - R_{1c} z_{1s}) (n_z n’_R - n_R n’_z + b_z b’_R - b_R b’_z)
-2 \tau\frac{G_0 \bar{B}}{B_0^2}.
\label{eq:Nhat}
\end{align}
In the last term of (\ref{eq:Nhat}), $\bar{B}$ is eliminated using (\ref{eq:const_B_on_axis}).
Applying the last two lines of (\ref{eq:Frenet}),
\begin{align}
n_z n’_R - n_R n’_z + b_z b’_R - b_R b’_z
=2 \tau R_0 
 + (n_R t_z-n_z t_R) \ell’ \kappa 
+n_z n_\phi + b_z b_\phi,
\end{align}
where we have used $n_z b_R-b_z n_R = t_\phi = R_0/\ell'$.
Applying
\begin{align}
n_z n_\phi + b_z b_\phi 
&=\vect{e}_z\cdot(\vect{t}\vect{t}+\vect{n}\vect{n}+\vect{b}\vect{b})\cdot\vect{e}_\phi-t_z t_\phi
=-t_z t_\phi = - R_0 z'_0 / (\ell')^2
\end{align}
and
\begin{align}
(n_z t_R -  n_R t_z )\kappa
= t_R \vect{e}_z \cdot \frac{d\vect{t}}{d\ell} - t_z \vect{e}_R \cdot\frac{d\vect{t}}{d\ell}
= \frac{R_0 z'_0 + R'_0 z''_0 - R''_0 z'_0}{(\ell')^3},
\end{align}
we find
\begin{align}
\hat{T}= \frac{|G_0|}{B_0 (\ell')^3}
(R_{1c} z_{1s} - R_{1s} z_{1c}) 
(R'_0 z''_0 + 2 R_0 z'_0 - z'_0 R''_0).
\label{eq:Nhat2}
\end{align}
Thus, $T^{FS}=T$ as desired.
This concludes the proof that whenever the curvature
of the magnetic axis does not vanish,
so the Frenet-Serret approach is free of singularities, all the equations derived directly in cylindrical coordinates in section \ref{sec:direct}
are equivalent to the analogous equations derived in 
the Frenet-Serret frame by \cite{GB1}.


\section{Quasisymmetry}
\label{sec:quasisymmetry}

Next, let us consider how the equations for the magnetic
field strength reduce in an important case, that of quasisymmetry.
(The more general condition of omnigenity will be considered
in Paper III.)
As shown by \cite{GB1}, for quasisymmetry to $O(r^1)$, the curvature of
the magnetic axis can never vanish, or else the elongation of the first-order flux surfaces diverges. Since the curvature does not vanish,
the Frenet-Serret frame is nonsingular, and the torsion
can be defined. Therefore the reduced equation
(\ref{eq:sigma_general}) should be free of singularities. We will consider the cases of quasi-axisymmetry and quasi-helical symmetry in turn.
We will not consider quasi-poloidal symmetry, $B=B(r,\theta)$,
since it cannot exist at $O(r^1)$.


\subsection{Quasi-axisymmetry}

Quasi-axisymmetry is the condition $\partial B/\partial\boozertor=0$. At $O(r^0)$, quasi-axisymmetry implies $B'_0=0$. It is convenient then to take the normalizing field $\bar{B}$ equal to the constant $\psisign B_0$, where $\psisign=\mathrm{sign}(\psi)=\pm 1$.
A consequence of $B'_0=0$ is $B_{1c}(\phi)=\hat{B}_{1c}(\boozertor_0)$ and $B_{1s}(\phi)=\hat{B}_{1s}(\boozertor_0)$.

At $O(r^1)$, quasi-axisymmetry implies $d \hat{B}_{1s}/d\boozertor=0$ and $d \hat{B}_{1c}/d\boozertor=0$. We are free to shift the origin of the $\theta$ coordinate so $\hat{B}_{1s}=0$, leaving the first-order magnetic field strength completely described by the single constant $\hat{B}_{1c}$. In this case, (\ref{eq:sigma_general}) simplifies to
\begin{align}
\frac{d\sigma}{d\boozertor}
+\iota_0 \left(
\frac{\hat{B}_{1c}^4}{B_0^4  \kappa^4}
+1+\sigma^2 \right)
-2 \left(\frac{I_2}{B_0} - \psisign\tau \right) \frac{ G_0 \hat{B}_{1c}^2}{ B_0^3 \kappa^2}=0,
\label{eq:QA}
\end{align}
where $\sigma(\boozertor)= \hat{B}_{1c} Y_{1c}(\boozertor) / \left(\Gsign \psisign B_0 \kappa(\boozertor)\right)$.
This result is equivalent to (82) in \cite{GB1} and to (A6) in \cite{GB2}.
In the appendix of Paper II \citep{PaperII}, we prove that for any given
$\sigma(0)$, $I_2 / B_0$, $G_0/B_0$, $\hat{B}_{1c}/(B_0\kappa)$, and $\psisign\tau$,
precisely one periodic solution $\sigma(\boozertor)$ and associated $\iota_0$ exist,
even though (\ref{eq:QA})  is nonlinear in $\sigma$.


\subsection{Quasi-helical symmetry}

Quasi-helical symmetry is the condition $B = B(r, M \theta - N \boozertor)$ for some nonzero integers $M$ and $N$. At $O(r^0)$,
this condition implies $B'_0=0$,
so again we can take $\bar{B}=\psisign B_0$ to normalize by the on-axis field. 
The fact that only $\propto \cos\theta$ and $\propto\sin\theta$ terms are permitted in first-order quantities like $\hat{B}_1$ means that  $M=1$ is required at this order. We are free to choose the origin of the $\theta$ coordinate so $\hat{B}_1(\theta,\boozertor)=\bar\eta B_0 \cos(\theta-N\boozertor)$ (for some constant $\bar\eta$), meaning $\hat{B}_{1c}=\bar\eta B_0 \cos(N\boozertor)$ and $\hat{B}_{1s} = \bar\eta B_0 \sin(N\boozertor)$. Substituting this $\hat{B}_{1s}$ and $\hat{B}_{1c}$ into (\ref{eq:sigma_general}), we find
\begin{align}
\frac{d\sigma}{d\boozertor}
+(\iota_0 - N) \left(
\frac{\bar\eta^4}{\kappa^4}
+1+\sigma^2 \right)
-2 \left(\frac{I_2}{B_0} - \psisign \tau \right) \frac{ G_0 \bar\eta^2}{ B_0 \kappa^2}=0.
\label{eq:QH}
\end{align}
Observe that (\ref{eq:QH}) is the same as the quasi-axisymmetry equation (\ref{eq:QA})
up to the generalizations $\hat{B}_{1c} \to \bar\eta B_0 $ and $\iota_0 \to \iota_0 - N$.
The same result can also be obtained by noting that if a helical angle $\vartheta = \theta - N \boozertor$ is introduced, 
(\ref{eq:straight_field_lines})-(\ref{eq:vacuum}) become
\begin{align}
\label{eq:theta_h}
\vect{B} = \nabla\psi\times\nabla\vartheta+(\iota-N)\nabla\boozertor\times\nabla\psi
=\beta\nabla\psi + I\nabla\vartheta + (G+NI)\nabla\boozertor.
\end{align}
These equations differ in form from (\ref{eq:straight_field_lines})-(\ref{eq:vacuum}) only through $\theta \to \vartheta$, $\iota \to \iota-N$, and $G\to G+NI$, with the latter replacement only having an effect at $O(r^2)$. Therefore, for $B$ to possess a single helicity in $\theta_h$ to the relevant order, the equations must be the same as for quasi-axisymmetry (in $\theta$) except for $\iota \to \iota-N$.

Furthermore, given a particular magnetic axis shape, it is possible to determine $N$ (including the quasi-axisymmetry case $N=0$)
before solving (\ref{eq:QA}) or (\ref{eq:QH}), by the following reasoning. Consider the general quasisymmetry condition  $\hat{B}_1(\theta,\boozertor) = \bar\eta B_0 \cos(\theta-N\boozertor)$ for constant $\bar\eta$,
where $N$ is allowed to be zero or nonzero, and let us take $\bar\eta>0$ without loss of generality.  Now consider a vector pointing perpendicularly from the axis to the $\theta-N\boozertor=0$ curve on the first-order-in-$r$ flux surface, which equivalently points to the maximum-$B$ contour on the surface. From (\ref{eq:GBr}) and (\ref{eq:GB_B1}), this vector is $\vect{n} r \bar\eta / \kappa + \vect{b} r Y_1$, which  has a positive projection along $\vect{n}$ at all $\boozertor$. Therefore this vector to the maximum-$B$ curve never points in a direction more than 90 degrees away from the normal vector $\vect{n}$. Hence, in a full toroidal transit around the axis, the $\theta-N\boozertor=0$ curve must wrap poloidally around the magnetic axis the same number of times $\vect{n}$ does so. Therefore, $N$ is the number of times $\vect{n}$ rotates poloidally around the axis in a full toroidal transit of the axis.
If $\vect{n}$ does not have such a net rotation for a given axis shape, then all quasisymmetric solutions for this axis shape will be quasi-axisymmetric, whereas if $\vect{n}$ does have this net rotation, all quasisymmetric solutions for this axis shape will be quasi-helically symmetric.

For another perspective on $N$, consider that because of (\ref{eq:theta_h}), in the derivation of (\ref{eq:sigma_general}), (\ref{eq:QA}), and (\ref{eq:QH}), it was never imposed that $\theta$ must be a poloidal angle rather than a helical angle. The choice of $N$ in the previous paragraph finally eliminates this redundancy. If one solves the quasi-axisymmetry equation (\ref{eq:QA}) for an axis shape that `really' should have quasi-helical symmetry rather than quasi-axisymmetry, one finds that the $\theta=0$ curve on each flux surface wraps around the axis poloidally as you traverse the axis toroidally, i.e. $\theta$ turns out to be a helical angle rather than a poloidal angle.

Numerical solution of (\ref{eq:QA})-(\ref{eq:QH}) as a practical
method to construct and parameterize quasisymmetric equilibria will be demonstrated in Paper II.


\subsection{Necessity of axis torsion}

Note that $\tau=0$ implies the magnetic axis and $\vect{n}$ are confined to a plane, so $\vect{n}$ cannot rotate poloidally about the magnetic axis. Then by the argument in the preceding section, $\tau=0$
can only be consistent with quasi-axisymmetry, not quasi-helical symmetry.
Moreover, in a stellarator, $I_2$ (which represents the on-axis density of toroidal current) is typically zero, as the bootstrap current vanishes on axis. In this case, if $\tau=0$, the integral of (\ref{eq:QA}) gives
\begin{align}
\iota_0 \int_0^{2\pi}d\boozertor \left[ \frac{\hat{B}_{1c}^4}{B_0^4 \kappa^4}+1+\sigma^2\right]=0.
\end{align}
The integral is positive-definite, so $\iota_0$ must vanish. 
Therefore, torsion of the magnetic axis is essential in a quasisymmetric stellarator in order to have rotational transform on axis.


\section{Discussion and conclusions}
\label{sec:conclusions}

In this paper, we have derived the relationship
near the magnetic axis
between the flux surface shape in cylindrical coordinates
and the magnetic field strength $B(r,\theta,\boozertor)$ in Boozer coordinates.
This relationship is important for stellarator design since $B(r,\theta,\boozertor)$ essentially determines the
guiding-center confinement, but it is the flux surface shape in three dimensions that determines the coils and engineering design.
As part of this calculation, we have also derived the relationship between the flux surface shape in cylindrical coordinates and the rotational transform.
No matter how low the aspect ratio of a stellarator,
the analysis here applies in a region sufficiently close to the axis.
The result of this analysis is the system of equations
(\ref{eq:const_B_on_axis}), (\ref{eq:B1s})-(\ref{eq:Kz}) or (\ref{eq:B1_mixedCoords}),
and (\ref{eq:iota})-(\ref{eq:denominator}). These equations can be derived directly in cylindrical coordinates, as in section \ref{sec:direct}, or by the appropriate transformation of Garren \& Boozer's equations, using the transformation of section \ref{sec:equivalence_transformation}. In contrast to the calculation of \cite{GB1}, the equations here
remain regular on segments or points where the axis torsion vanishes, which always occurs for omnigenous fields with poloidally closed $B$ contours. The torsion, which may not be well defined in this circumstance, does not appear in our analysis since we avoid using the Frenet-Serret frame.

Consistent with \cite{GB1}, we find that at $O(r^1)$,
for a prescribed $B_1$, there are two more $\phi$-dependent
degrees of freedom than there are equations.
Specifically, the six $\phi$-dependent unknowns ($R_0$, $z_0$, 
$R_{1c}$, $R_{1s}$, $z_{1c}$, and $z_{1s}$) are constrained
by four equations: (\ref{eq:const_B_on_axis}), the $\sin\theta$ and $\cos\theta$ components of 
(\ref{eq:B1s}), and (\ref{eq:iota}). Thus, two of these six functions can be viewed as inputs. Choosing $R_0$ and $z_0$ as the two inputs amounts to specifying the magnetic axis shape, and the four aforementioned equations then give the flux surface shape that yields the desired $B_1$.


We acknowledge illuminating conversations about this work with 
Gabriel Plunk.
This work was supported by the Simons Foundation and by the
U.S. Department of Energy, Office of Science, Office of Fusion Energy Science,
under award numbers DE-FG02-93ER54197 and DE-FG02-86ER53223.
This work was also supported by a grant from the Simons Foundation (560651, ML).


\appendix

\section{\changed{Regularity near the magnetic axis}}
\label{sec:regularity}

\changed{
In this section we will derive the form of the expansion (\ref{eq:RExpansion})-(\ref{eq:R2Expansion}) for $R$, $z$, $\nu$, and $B$. As an alternative to the argument based on analyticity in \cite{GB1}, here we give a constructive demonstration, proceeding in several steps. First, we will derive the form (\ref{eq:RExpansion})-(\ref{eq:R2Expansion}) for $R$ and $z$  but with a non-straight-field-line poloidal angle $\alpha$  in place of the Boozer angle $\theta$. Then we will derive the form (\ref{eq:RExpansion})-(\ref{eq:R2Expansion}) for $R$ and $z$ but with the poloidal angle $\xi$ defined such that field lines are straight in the $\xi$-$\phi$ plane. Next, we will derive (\ref{eq:RExpansion})-(\ref{eq:R2Expansion}) for $\theta$. Finally, we extend the proof to $\nu$  and $B$.
}

\changed{
Assuming good flux surfaces exist near the axis, a Taylor expansion exists for $\psi(R,z)$:
\begin{align}
\label{eq:psi_Taylor}
\psi = &\frac{\dR^2}{2}\psi_{RR} + \dR \dz \psi_{Rz} + \frac{\dz^2}{2}\psi_{zz} 
+\frac{\dR^3}{6}\psi_{RRR}
\\ &
+\frac{\dR^2 \dz}{2} \psi_{RRz}
+\frac{\dR \dz^2}{2} \psi_{Rzz}
+\frac{\dz^3}{6}\psi_{zzz} + \ldots ,
\nonumber
\end{align}
where quantities such as $\psi_{RR}$ 
refer to partial derivatives evaluated at the axis $(R_0,z_0)$,
and dependence on the independent variable $\phi$ is not displayed to simplify notation.
Note $A>0$ where $A=\psi_{RR}\psi_{zz} - \psi_{Rz}^2$, since the axis is an extremum of $\psi$ rather than a saddle point. For this section we assume $\psi_{RR}$ and $\psi_{zz}$ are positive for simplicity, so $\psi\ge 0$.
We then seek a solution of the desired form:
\begin{align}
\label{eq:RzExpansions}
R = &R_0 + r (R_{1c}^\alpha \cos\alpha + R_{1s}^\alpha \sin\alpha ) + r^2 (R_{20}^\alpha + R_{2c}^\alpha \cos 2\alpha + R_{2s}^\alpha \sin 2\alpha) + O(r^3),
\\
z = & z_0 + r (z_{1c}^\alpha \cos\alpha + z_{1s}^\alpha \sin\alpha) + r^2(z_{20}^\alpha + z_{2c}^\alpha \cos 2\alpha + z_{2s}^\alpha \sin 2\alpha) + O(r^3).
\nonumber
\end{align}
Substituting (\ref{eq:RzExpansions}) into (\ref{eq:psi_Taylor}), terms can be collected based on their order in $r$ and $\alpha$ dependence. The number of equations that result at a given order in $r$ is smaller than the number of associated coefficients in (\ref{eq:RzExpansions}), reflecting the non-uniqueness of the poloidal angle; for instance the $\alpha=0$ direction can be shifted. 
One solution satisfying (\ref{eq:psi_Taylor}) through $O(r^3)$ is $z_{1c}^\alpha=0$, $R_{20}^\alpha=0$,
\begin{align}
\label{eq:alpha_solution}
R_{1c}^\alpha = &\sqrt{\frac{\bar{B}}{\psi_{RR}}}
,\hspace{0.2in}
R_{1s}^\alpha = \psi_{Rz}\sqrt{\frac{\bar{B}}{\psi_{RR} A}}
,\hspace{0.2in}
z_{1s}^\alpha = -\sqrt{\frac{\bar{B} \psi_{RR}}{A}}
,\hspace{0.2in}
R_{2c}^\alpha= 
-\frac{\bar{B} \psi_{RRR}}{6 \psi_{RR}^2},
\\
R_{2s}^\alpha= &
-\frac{\bar{B}}{12 \psi_{RR}^2 A^{5/2}} \left[
\psi_{RRR} \left(4 \psi_{RR}^2 
\psi_{Rz} \psi_{zz}^2-5 \psi_{RR} \psi_{Rz}^3 \psi_{zz}+2
   \psi_{Rz}^5\right) \right. 
\nonumber   \\
   & \hspace{0.9in} \left.-\psi_{RR}^3 \left(3 \psi_{RRz} \psi_{zz}^2+\psi_{Rz}^2 \psi_{zzz}-3 \psi_{Rz}
   \psi_{Rzz} \psi_{zz}\right)\right],
   \nonumber \\
z_{2c}^\alpha= & -z_{20}^\alpha=
\frac{\bar{B}}{12 \psi_{RR} A^2}
\left[\psi_{RR}^3 \psi_{zzz}-3 \psi_{RR}^2 \psi_{Rz} \psi_{Rzz}+3 \psi_{RR} \psi_{RRz} \psi_{Rz}^2-\psi_{RRR} \psi_{Rz}^3\right],
\nonumber \\
z_{2s}^\alpha= &-
\frac{\bar{B}}{12 \psi_{RR} A^{5/2}} \left[
\psi_{RR}^3 (\psi_{Rz} \psi_{zzz}-3 \psi_{Rzz} \psi_{zz})+\psi_{RR}^2 \psi_{zz} (6 \psi_{RRz} \psi_{Rz}-\psi_{RRR} \psi_{zz})
\right. 
\nonumber   \\
   & \hspace{0.9in} \left.-\psi_{RR} \psi_{Rz}^2 (\psi_{RRR} \psi_{zz}+3 \psi_{RRz} \psi_{Rz})+\psi_{RRR} \psi_{Rz}^4\right].
   \nonumber
\end{align}
Thus, given the Taylor series for $\psi(R,z)$, we can  construct expansions of the form (\ref{eq:RExpansion})-(\ref{eq:R2Expansion}), but with $\theta \to \alpha$, for $R$ and $z$.
The $O(r)$ terms in (\ref{eq:RzExpansions}) can be manipulated to write the poloidal angle explicitly as
\begin{align}
\alpha \approx \mathrm{atan2}\left( -\dz, \; 
[\dR \psi_{RR} +\dz \psi_{Rz}]/\sqrt{A} \right),
\label{eq:alpha_def}
\end{align}
where atan2 is the arctangent with range $(-\pi, \; \pi]$.
}

\changed{
Next we construct the straight-field-line angle $\xi = \alpha + \lambda$ where $\lambda(r,\alpha,\phi)$ is single-valued. From the $\nabla\phi$ component of $\vect{B}=\nabla \psi \times \nabla \xi + \iota\nabla\phi\times\nabla\psi$, we find
\begin{align}
\label{eq:lambda_def}
\lambda = f(r,\phi) + \int_0^\alpha d\alpha' \left[ 
\left( \frac{\partial \vect{r}}{\partial\psi}\cdot\frac{\partial\vect{r}}{\partial\alpha} \times \frac{\partial\vect{r}}{\partial\phi}\right)\vect{B}\cdot\nabla\phi
- 1\right],
\end{align}
for some $f(r,\phi)$. 
The Jacobian in this expression can be evaluated using derivatives of $\vect{r}=R\vect{e}_R + z \vect{e}_z$; substitution of (\ref{eq:RzExpansions}) then yields
\begin{align}
\frac{\partial \vect{r}}{\partial\psi}\cdot\frac{\partial\vect{r}}{\partial\alpha} \times \frac{\partial\vect{r}}{\partial\phi}
=\left( \frac{\partial z}{\partial \psi}\frac{\partial R}{\partial \alpha} - \frac{\partial z}{\partial \alpha}\frac{\partial R}{\partial \psi} \right) R
= \frac{R_0}{\sqrt{A}}\left( 1 + r J_{1s} \sin\alpha + r J_{1c} \cos\alpha + O(r^2)\right),
\end{align}
where $J_{1s}$ and $J_{1c}$ are complicated algebraic functions of the Taylor coefficients in (\ref{eq:psi_Taylor}).
Also, in (\ref{eq:lambda_def}), $\vect{B}\cdot\nabla\phi$ is smooth so it has a Taylor series
\begin{equation}
\label{eq:Bphi_Taylor}
\vect{B}\cdot\nabla\phi
=b_0 + \dR b_R + \dz b_z + O(r^2).
\end{equation}
Using the $O(r)$ terms in (\ref{eq:RzExpansions}) and (\ref{eq:alpha_solution}) in eq (\ref{eq:ellipse_area}) for the area of an ellipse, flux surfaces near the axis have an area $2\pi\psi/\sqrt{A}$ in the $R$-$z$ plane, so $b_0 = \sqrt{A}/R_0$. Evaluating the integral in (\ref{eq:lambda_def}) then gives
\begin{align}
\label{eq:lambda_solution}
\lambda = \hat{f}(r,\phi) + r \lambda_{1s}\sin\alpha + r \lambda_{1c} \cos\alpha + O(r^2),
\end{align}
where $\lambda_{1s}=b_R R_{1c}^\alpha R_0 / \sqrt{A} + J_{1c}$,
$\lambda_{1c} = -J_{1s}-(b_R R_{1s}^\alpha+b_z z_{1s}^\alpha) R_0 / \sqrt{A}$, and $\hat{f}=f-r\lambda_{1c}$.
To constrain the form of $\hat{f}$, we use the $\nabla\alpha$ component of $\vect{B}=\nabla\psi\times\nabla\xi+\iota\nabla\phi\times\nabla\psi$ to write
\begin{align}
\label{eq:d_lambda_d_phi}
\frac{\partial\lambda}{\partial\phi}
=\iota - \frac{\vect{B}\cdot\nabla\alpha}{\nabla\psi\cdot\nabla\alpha\times\nabla\phi}
=\iota - \vect{B}\cdot\frac{\partial\vect{r}}{\partial\phi}\times\frac{\partial \vect{r}}{\partial\psi}.
\end{align}
In the last term, note that $\vect{B}$ has a Taylor expansion in $R-R_0$ and $z-z_0$ like (\ref{eq:Bphi_Taylor}) but with vector coefficients; the leading term is parallel to $\partial\vect{r}/\partial \phi$, so the last term in (\ref{eq:d_lambda_d_phi}) is finite on the axis.
Evaluating the last term in (\ref{eq:d_lambda_d_phi}) by differentiating $\vect{r}=R\vect{e}_R+z\vect{e}_z$ and substituting (\ref{eq:RzExpansions}), and applying $\int_0^{2\pi} d\alpha \; \partial(\ldots)/\partial r$ to (\ref{eq:d_lambda_d_phi}), we find $\int_0^{2\pi} d\alpha \; \partial^2 \lambda /\partial r \partial \phi=0$ at $r=0$, which implies
the $O(r)$ term of $\hat{f}$ is independent of $\phi$. This term can therefore be set to 0, since $\lambda$ can be shifted by any function of only $r$. Hence,
\begin{align}
\label{eq:lambda_solution2}
\lambda = \lambda_0 + r \lambda_{1s}\sin\alpha + r \lambda_{1c} \cos\alpha + O(r^2),
\end{align}
for some $\lambda_0(\phi)$.
Substituting $\alpha = \xi - \lambda$ and (\ref{eq:lambda_solution2}) into (\ref{eq:RzExpansions}), we obtain an expansion of the desired form:
\begin{align}
\label{eq:RzExpansions_xi}
R = &R_0 + r (R_{1c}^\xi \cos\xi + R_{1s}^\xi \sin\xi ) + r^2 (R_{20}^\xi + R_{2c}^\xi \cos 2\xi + R_{2s}^\xi \sin 2\xi) + O(r^3),
\\
z = & z_0 + r (z_{1c}^\xi \cos\xi + z_{1s}^\xi \sin\xi) + r^2(z_{20}^\xi + z_{2c}^\xi \cos 2\xi + z_{2s}^\xi \sin 2\xi) + O(r^3),
\nonumber
\end{align}
where the $R^\xi$ and $z^\xi$ coefficients are functions of the $R^\alpha$ and $z^\alpha$ coefficients, e.g. 
$R_{1s}^\xi = R_{1s}^\alpha \cos\lambda_0 + R_{1c}^\alpha\sin\lambda_0$.
}

\changed{
Next we transform to Boozer coordinates. The magnetic field can be written \citep{HelanderReview} as
\begin{align}
\label{eq:B_PEST}
\vect{B} = \hat{\beta}\nabla\psi + I \nabla \xi + G \nabla \phi + \nabla [(G+\iota I)\nu],
\end{align}
for some $\hat{\beta}$, where the transformation to Boozer coordinates is given by $\varphi = \phi + \nu$ and $\theta = \xi + \iota \nu$. Applying $\nabla\phi \times \nabla\psi \cdot ( \ldots)$ to (\ref{eq:B_PEST}), we find
\begin{align}
\label{eq:nu_integral}
\nu = g(r,\phi) + \frac{1}{G+\iota I}
\int_0^{\xi} d\xi'
\left[ \frac{\vect{B}\cdot\nabla\phi\times\nabla\psi}{\vect{B}\cdot\nabla\phi} - I\right].
\end{align}
The denominator is smooth (and nonvanishing near the axis for cases of interest in this paper), with the expansion (\ref{eq:Bphi_Taylor}). The numerator is a product of three  quantities that are smooth near the axis and so it too is  smooth, vanishing on the axis since $\nabla\psi=0$ there. Noting $I$ is smooth function of $\psi$ and $I=0$ on axis, 
then the quantity in square brackets in (\ref{eq:nu_integral}) is smooth and so has a Taylor expansion
\begin{align}
 \frac{\vect{B}\cdot\nabla\phi\times\nabla\psi}{\vect{B}\cdot\nabla\phi} - I
=&\dR H_R + \dz H_z  
+ \frac{1}{2} \dR^2 H_{RR} \\
&+ \dR \dz H_{Rz} + \frac{1}{2} \dz^2 H_{zz} + O(r^3),
\nonumber
\end{align}
for some coefficients $H_{\ldots}$. Substituting (\ref{eq:RzExpansions_xi}) and integrating in $\xi$,
(\ref{eq:nu_integral}) gives
\begin{align}
\nu = \hat{g}(r,\phi) + 
r (\nu_{1s}^\xi \sin\xi + \nu_{1c}^\xi \cos\xi)
+r^2 (\nu_{20}^\xi+\nu_{2s}^\xi\sin 2\xi + \nu_{2c}^\xi\cos 2\xi)+O(r^3),
\end{align}
where $\hat{g}$ is the sum of $g$ and terms from the lower integration bound.
To constrain the form of $\hat{g}$ we apply $\nabla\psi\times\nabla\xi\cdot(\ldots)$ to (\ref{eq:B_PEST}), with the result
\begin{align}
\frac{\partial\nu}{\partial\phi}
=\frac{1}{G+\iota I}
\left[ \frac{B^2 - \iota \vect{B}\cdot\nabla\phi\times\nabla\psi}{\vect{B}\cdot\nabla\phi}-G\right].
\end{align}
The right-hand side is manifestly smooth near the axis and so it has a Taylor series in $R$ and $z$, into which we substitute (\ref{eq:RzExpansions_xi}). Applying $\partial/\partial r$ and integrating over $\xi$, we find $\int_0^{2\pi} d\xi \; \partial^2 \nu/\partial r\partial\phi = 0$ at $r=0$. It follows that the $\partial \hat{g}/\partial\phi$ has no term linear in $r$. Then since we are free to shift $\nu$ by any function of only $r$, we can choose $\hat{g}$ so $\nu$ has the form
\begin{align}
\label{eq:nu_xi}
\nu = \nu_0(\phi) + 
r (\nu_{1s}^\xi \sin\xi + \nu_{1c}^\xi \cos\xi)
+r^2 (\nu_{20}^\xi+\nu_{2s}^\xi\sin 2\xi + \nu_{2c}^\xi\cos 2\xi)+O(r^3).
\end{align}
Substitution of $\xi = \theta - \iota \nu$ and (\ref{eq:nu_xi}) in (\ref{eq:RzExpansions_xi}) yields
the desired expansions for $R$ and $z$, (\ref{eq:RExpansion})-(\ref{eq:R2Expansion}). The same substitutions applied to
(\ref{eq:nu_xi}) give the desired expansion for $\nu(r,\theta,\phi)$.
}

\changed{
Finally, $B$ is smooth near the axis and so it has a Taylor expansion
\begin{align}
\label{eq:B_Taylor_series}
 B
=&B_0 + \dR B_R + \dz B_z  
+ \frac{1}{2} \dR^2 B_{RR} \\
&+
\dR \dz B_{Rz} + \frac{1}{2} \dz^2 B_{zz} + O(r^3).
\nonumber
\end{align}
Substitution of (\ref{eq:RExpansion})-(\ref{eq:R2Expansion})
for $R(r,\theta,\phi)$ and the analogous expansion for $z(r,\theta,\phi)$ into (\ref{eq:B_Taylor_series})
gives the desired expansion for $B(r,\theta,\phi)$.
}


\section{Geometric properties of flux surfaces}
\label{sec:ellipse}

\begin{figure}
  \centering
  \includegraphics[width=2in]{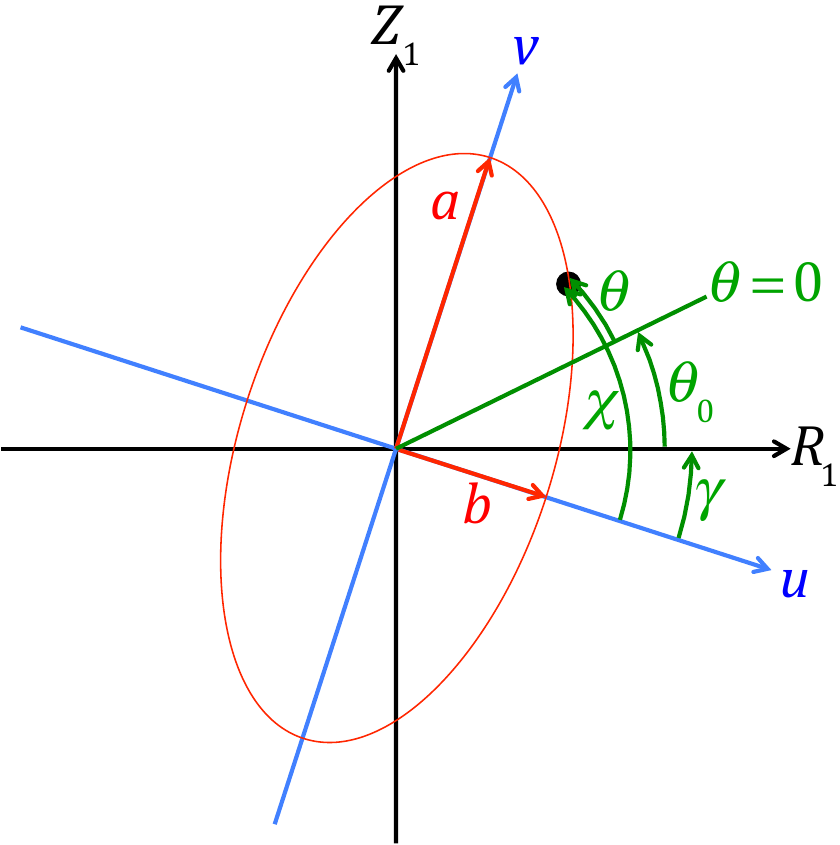}
  \caption{
  Definitions for appendix \ref{sec:ellipse}.
  }
\label{fig:ellipse}
\end{figure}

Here we relate several geometric properties of the flux surfaces -- specifically the cross-sectional area and elongation -- to the variables $(R_1,z_1)$ used elsewhere in the paper.
We consider a cross section of the flux surfaces in a constant-$\phi$ plane. All results of this section apply to cross sections perpendicular to the magnetic axis if $(R_1,z_1)$ are replaced by $(X_1,Y_1)$.
Several geometric quantities are defined in figure \ref{fig:ellipse}.
To $O(r)$, the flux surfaces are elliptical, with semi-major axis $a$ and semi-minor axis $b$. Axes $u$ and $v$ are aligned with the minor and major axes, and $\gamma$ is the angle between the $u$ and $R_1$ axes. 
The $\theta=0$ line is not generally aligned with any of these axes,
and we let $\theta_0$ denote the angle between this line and the $R_1$ axis.
Any point
in the plane, such as the black dot in the figure, makes an
angle $\theta+\theta_0$ relative to the $R_1$ axis and an angle $\chi$
relative to the $u$ axis, with $\chi = \theta +\theta_0+ \gamma$.
Substituting $u = b\cos\chi$ and $v = a\sin\chi$
into
\begin{align}
\begin{pmatrix} R_1\\z_1 \end{pmatrix}
=
\begin{pmatrix} \cos\gamma & \sin\gamma \\
-\sin\gamma & \cos\gamma \end{pmatrix}
\begin{pmatrix} u\\v \end{pmatrix},
\end{align}
applying the angle sum formula to $\chi$,
and equating $\sin\theta$ and $\cos\theta$ terms using
(\ref{eq:R1}), we find
\begin{align}
\label{eq:Rz_to_ellipse}
\begin{pmatrix} R_{1s} \\ R_{1c} \end{pmatrix}
=
\begin{pmatrix} \cos\theta_0 & -\sin\theta_0 \\
\sin\theta_0 & \cos\theta_0 \end{pmatrix}
\begin{pmatrix} (a-b)\sin\gamma \cos\gamma\\
a \sin^2\gamma + b\cos^2\gamma \end{pmatrix},
\\
\begin{pmatrix} z_{1s} \\ z_{1c} \end{pmatrix}
=
\begin{pmatrix} \cos\theta_0 & -\sin\theta_0 \\
\sin\theta_0 & \cos\theta_0 \end{pmatrix}
\begin{pmatrix} a \cos^2\gamma + b\sin^2\gamma\\
(a-b)\sin\gamma \cos\gamma\end{pmatrix}.
\nonumber
\end{align}
Using (\ref{eq:Rz_to_ellipse}), the right-hand side of (\ref{eq:const_B_on_axis})
is found to be
\begin{align}
\label{eq:ellipse_area}
R_{1s}z_{1c}-R_{1c}z_{1s} = -ab,
\end{align}
which is (minus) the area of the ellipse divided by $\pi$.

Another important property of the flux surfaces is their elongation, $a/b$. 
In practice, many solutions of equation (\ref{eq:sigma_general}) are uninteresting since they correspond to impractically large values of elongation, so to discard these solutions it is valuable to derive an expression for the elongation in terms of $R_1$ and $z_1$.
Such a formula can be obtained by first 
defining $p=R_{1s}^2 + R_{1c}^2 + z_{1s}^2 + z_{1c}^2$,
and noting from (\ref{eq:Rz_to_ellipse}) that
$p=a^2+b^2$.
Then defining $q=R_{1s}z_{1c}-R_{1c}z_{1s} = -ab$, we can solve
$a^4- p a^2 + q^2 = 0$ for $a$, noting the larger positive root is $a$ and the smaller is $b$, since $b$ satisfies the same quadratic equation. Then the elongation is
\begin{equation}
\frac{a}{b} = \sqrt{\frac{p + \sqrt{p^2-4 q^2}}{p-\sqrt{p^2-4 q^2}}}
=\frac{p+\sqrt{p^2-4q^2}}{2|q|}.
\end{equation}


\section{Equating representations of the field: second order}
\label{sec:2nd_order}

Here the derivation of (\ref{eq:iota})-(\ref{eq:denominator}) is presented.
The $O(r^2)$ terms in 
(\ref{eq:BR})-(\ref{eq:Bz}) 
can be obtained by applying $\partial/\partial r$ twice and evaluating
the results at $r\to 0$. We find
\begin{align}
\label{eq:BR2}
\frac{\bar{B}}{G_0 R_0} \left[
\frac{\partial R_1}{\partial\phi} + \iota_0 \left(1+\nu'_0\right) \frac{\partial R_1}{\partial\theta}
-\frac{R_1}{R_0} R'_0-\iota_0 \frac{\partial \nu_1}{\partial\theta} R'_0
+\beta_0 R_0\frac{\partial z_1}{\partial\theta}\right]
\\
= \frac{I_2 z_1}{G_0}
+\frac{\partial \nu_2}{\partial\theta} z_1 + 2 \frac{\partial \nu_1}{\partial\theta} z_2 
- \nu_1 \frac{\partial z_2}{\partial\theta} - 2 \nu_2 \frac{\partial z_1}{\partial\theta},
\nonumber
\end{align}
\begin{align}
\label{eq:2nd_order_phi}
&\frac{\bar{B}}{G_0 R_0} \left[
-\frac{R_1}{R_0} \left(\ell'\right)^2 - \iota_0 \frac{\partial\nu_1}{\partial\theta} \left(\ell'\right)^2
+2 R_0 R_1 + 2 R'_0 \frac{\partial R_1}{\partial\phi} + 2 z'_0\frac{\partial z_1}{\partial\phi} 
\right. \\
& \left.\hspace{0.5in} +\iota_0\left( 1 + \frac{d\nu_0}{d\phi}\right) \left( \frac{\partial R_1}{\partial\theta} R'_0 + \frac{\partial z_1}{\partial\theta} z'_0\right)\right]
\nonumber \\
&=\left( 2 z_2 \frac{\partial R_1}{\partial\theta} + z_1 \frac{\partial R_2}{\partial\theta} - 2 R_2 \frac{\partial z_1}{\partial\theta} - R_1 \frac{\partial z_2}{\partial\theta}\right)
\left( 1 + \nu'_0\right)
+\left( z_1 \frac{\partial R_1}{\partial\theta} - R_1 \frac{\partial z_1}{\partial\theta}\right) \frac{\partial\nu_1}{\partial\phi},
\nonumber 
\end{align}
\begin{align}
\label{eq:Bz2}
\frac{\bar{B}}{G_0 R_0} \left[
\frac{\partial z_1}{\partial\phi} + \iota_0 \left(1+\nu'_0\right) \frac{\partial z_1}{\partial\theta}
-\frac{R_1}{R_0} z'_0-\iota_0 \frac{\partial \nu_1}{\partial\theta} z'_0
-\beta_0 R_0 \frac{\partial R_1}{\partial\theta}
\right]
 \\
=-\frac{I_2 R_1}{G_0}
+\frac{\partial R_2}{\partial\theta} \nu_1 + 2 \frac{\partial R_1}{\partial\theta} \nu_2 
- R_1 \frac{\partial \nu_2}{\partial\theta} - 2 R_2 \frac{\partial \nu_1}{\partial\theta}.
\nonumber
\end{align}
In (\ref{eq:2nd_order_phi}), the terms including a factor of $\iota_0$ can be written in the combination
(\ref{eq:iota_terms}), which vanishes as before.
Plugging in (\ref{eq:R1})-(\ref{eq:R2Expansion}), it can be seen that
 (\ref{eq:BR2})-(\ref{eq:Bz2}) each have only $\sin\theta$ and $\cos\theta$ Fourier components. These $\sin\theta$ and $\cos\theta$ components give the following six equations:
\begin{align}
\label{eq:second_order_Rs}
&\frac{\bar{B}}{2 G_0 R_0} \left[
R'_{1s} 
- \iota_0 \left(1+\nu'_0\right) R_{1c}
-\frac{R_{1s}}{R_0} R'_0+\iota_0 \nu_{1c} R'_0
-\beta_0 R_0 z_{1c} \right] \\
&= \frac{I_2 z_{1s}}{2G_0}
+\nu_{1c}\left( z_{2c}-z_{20}\right) + \nu_{1s} z_{2s} + z_{1c} \left( \nu_{20}-\nu_{2c}\right) - z_{1s} \nu_{2s},  \nonumber
\end{align}
\begin{align}
\label{eq:second_order_Rc}
&\frac{\bar{B}}{2 G_0 R_0} \left[
R'_{1c}
+ \iota_0 \left(1+\nu'_0\right) R_{1s} -\frac{R_{1c}}{R_0} R'_0-\iota_0 \nu_{1s} R'_0
+\beta_0 R_0 z_{1s}\right]  \\
&= \frac{I_2 z_{1c}}{2G_0}
+\nu_{1s} \left(z_{2c}+z_{20}\right) - \nu_{1c} z_{2s} - z_{1s} \left( \nu_{20} + \nu_{2c}\right) + z_{1c} \nu_{2s}, \nonumber
\end{align}
\begin{align}
\label{eq:second_order_zs}
&\frac{\bar{B}}{2G_0 R_0} \left[
z'_{1s}
- \iota_0 \left(1+\nu'_0\right) z_{1c}
-\frac{R_{1s}}{R_0} z'_0+\iota_0 \nu_{1c} z'_0
+\beta_0 R_0 R_{1c} \right]  \\
&=-\frac{I_2 R_{1s}}{2G_0}
+
\nu_{1c}\left( R_{20} - R_{2c}\right) - \nu_{1s} R_{2s} + R_{1c} \left( \nu_{2c}-\nu_{20}\right) + R_{1s} \nu_{2s}, \nonumber
\end{align}
\begin{align}
\label{eq:second_order_zc}
&\frac{\bar{B}}{2 G_0 R_0} \left[
z'_{1c}
+ \iota_0 \left(1+\nu'_0\right) z_{1s} -\frac{R_{1c}}{R_0} z'_0-\iota_0 \nu_{1s} z'_0
-\beta_0 R_0 R_{1s}\right]  \\
&=-\frac{I_2 R_{c1}}{2G_0}
-\nu_{1s} \left( R_{20} + R_{2c}\right) + \nu_{1c} R_{2s} + R_{1s} \left( \nu_{2c} + \nu_{20}\right) - R_{1c} \nu_{2s}, \nonumber
\end{align}
\begin{align}
\label{eq:second_order_phis}
&\frac{\bar{B}}{G_0 R_0} \left[
-\frac{R_{1s}}{2R_0} \left(\ell'\right)^2 
+ R_0 R_{1s} + R'_0 
R'_{1s}
+ z'_0
z'_{1s}
\right]
 \\
&=\left[
z_{1c} \left(R_{20} - R_{2c}\right) - z_{1s} R_{2s} + R_{1c}\left( z_{2c} - z_{20}\right) + R_{1s} z_{2s}
\right]
\left( 1 + \nu'_0\right)
+\frac{\Gsign \bar{B}}{2 R_0 B_0}\ell' 
\nu'_{1s},
\nonumber
\end{align}
\begin{align}
\label{eq:second_order_phic}
&\frac{\bar{B}}{G_0 R_0} \left[
-\frac{R_{1c}}{2R_0} \left(\ell'\right)^2 
+ R_0 R_{1c} + R'_0 
R'_{1c}
+ z'_0
z'_{1c}
\right]
\\
&=\left[
-z_{1s}\left( R_{2c} + R_{20}\right) + z_{1c} R_{2s} + R_{1s} \left( z_{20} + z_{2c}\right) - R_{1c} z_{2s}
\right]
\left( 1 + \nu'_0\right)
+\frac{\Gsign \bar{B}}{2 R_0 B_0}\ell' 
\nu'_{1c}.
\nonumber 
\end{align}
In the last two equations we have used (\ref{eq:const_B_on_axis}). 

While these six equations contain $R_2$, $\nu_2$, and $z_2$, all these subscript-2 quantities can be eliminated to give a constraint on the subscript-1 quantities by forming
\begin{equation}
\label{eq:annihilator}
\left( 1 + \nu'_0\right)
\left[  
\mathrm{(\ref{eq:second_order_Rs})} R_{1c} -  \mathrm{(\ref{eq:second_order_Rc})} R_{1s} 
+\mathrm{(\ref{eq:second_order_zs})} z_{1c} -  \mathrm{(\ref{eq:second_order_zc})} z_{1s} 
\right]
-\mathrm{(\ref{eq:second_order_phis})} \nu_{1c} +  \mathrm{(\ref{eq:second_order_phic})} \nu_{1s} .
\end{equation}
The $\beta_0$ terms happen to vanish as well in this combination.
Multiplying the result through by $2 G_0 R_0/\bar{B}$, we obtain
\begin{align}
&\left( 1 + \nu'_0\right)\left[ 
R_{1c} R'_{1s} - R_{1s} R'_{1c}
+z_{1c} z'_{1s} - z_{1s} z'_{1c}
+(R_{1c} z_{1s} - R_{1s} z_{1c}) z'_0 / R_0 \right. \nonumber \\
&\hspace{0.8in}
-\iota_0 \left( 1 + \nu'_0\right)
\left( R_{1c}^2 + R_{1s}^2 + z_{1c}^2 + z_{1s}^2 \right)
\nonumber \\
&\hspace{0.8in}\left.
+ \iota_0 R'_0 \left( \nu_{1c} R_{1c} + \nu_{1s} R_{1s} \right)
+ \iota_0 z'_0 \left( \nu_{1c} z_{1c} + \nu_{1s} z_{1s} \right)
\right] \nonumber \\
&- 2 \nu_{1c} \left[
-\frac{R_{1s}}{2R_0} \left(\ell'\right)^2 
+ R_0 R_{1s} + R'_0 R'_{1s} + z'_0 z'_{1s}
\right] \nonumber \\
&+ 2 \nu_{1s}  \left[
-\frac{R_{1c}}{2R_0} \left(\ell'\right)^2 
+ R_0 R_{1c} + R'_0 R'_{1c} + z'_0 z'_{1c}
\right] \nonumber \\
&= \frac{|G_0|}{B_0} \ell' \left( \nu_{1s} \nu'_{1c} - \nu_{1c} \nu'_{1s} \right)
+\frac{2 I_2 R_0}{\bar{B}}\left(1+\nu'_0 \right)
\left(R_{1c}z_{1s}-R_{1s}z_{1c} \right).
\end{align}

Eliminating $\nu_0$, we find
$(T - \iota_0 V)(\ell')^2 B_0^2/G_0^2=0$
where 
\begin{align}
T = &
\frac{|G_0|^3}{B_0^3 \ell'}  \left( 
\nu_{1c} \nu'_{1s} - \nu_{1s} \nu'_{1c} \right)
\\
&+\frac{|G_0|}{B_0 \ell'}
 \left[ 
R_{1c} R'_{1s} - R_{1s} R'_{1c}
+z_{1c} z'_{1s} - z_{1s} z'_{1c}
+\frac{(R_{1c} z_{1s} - R_{1s} z_{1c})}{R_0} z'_0 \right] \nonumber \\
&-  \frac{2 G_0^2 \nu_{1c}}{B_0^2(\ell')^2}
 \left[
-\frac{R_{1s}}{2R_0} \left(\ell'\right)^2 
+ R_0 R_{1s} + R'_0 R'_{1s} + z'_0 z'_{1s}
\right] \nonumber \\
&+  \frac{2 G_0^2 \nu_{1s} }{B_0^2(\ell')^2}
 \left[
-\frac{R_{1c}}{2R_0} \left(\ell'\right)^2 
+ R_0 R_{1c} + R'_0 R'_{1c} + z'_0 z'_{1c}
\right] 
+\frac{2 I_2 G_0}{B_0^2} , \nonumber
\end{align}
and 
\begin{align}
V&=
 R_{1c}^2 + R_{1s}^2 + z_{1c}^2 + z_{1s}^2  - \frac{|G_0|}{B_0 \ell'} \left[R'_0 \left( \nu_{1c} R_{1c} + \nu_{1s} R_{1s} \right)
+ z'_0 \left( \nu_{1c} z_{1c} + \nu_{1s} z_{1s} \right) \right].
\end{align}
Eliminating $\nu_{1s}$ and $\nu_{1c}$ using (\ref{eq:nu1})
results in (\ref{eq:numerator})-(\ref{eq:denominator}).

\bibliographystyle{jpp}

\bibliography{quasisymmetry_cylindrical_I}

\end{document}